\documentclass[11pt]{article}\topmargin=-0.7cm \textheight 222mm \textwidth 15.8cm\oddsidemargin=4mm 
\newcommand{\comm}[1]{}

\def\xxxonly{ }
 \usepackage{color}

\usepackage[numbers]{natbib}\usepackage{graphicx} \usepackage{epsfig}\def\citet{\cite}
\def\citet{\cite}
\usepackage{amssymb}
\usepackage{setspace}
\usepackage{ae} 
\usepackage{eucal} 

\newtheorem{theorem}{Theorem}
\newtheorem{proposition}{Proposition}

\newtheorem{definition}{Definition}
\newtheorem{remark}{Remark}

\def\e{\varepsilon}

\def\defi{\stackrel{{\scriptscriptstyle \Delta}}{=}}

\def\a{\alpha}
\def\d{\delta}
\def\o{\omega}

\def\Y{{\cal Y}}

\def\w{\widehat}
\def\Ind{{\mathbb{I}}}

\def\sign{{\rm  sign\,}}

\def\R{{\bf R}}

\def\Z{{\cal Z}}

\def\g{\gamma}
\def\C{{\bf C}}

\def\ww{\widetilde}
\def\X{{\cal X}}
\def\t{\theta}
\def\oo{\bar}

 \def\V{{\cal V}}

\def\M{{\cal M}}
\def\B{{\cal B}}

\newcommand{\be}{\begin{equation}}
\newcommand{\ee}{\end{equation}}
\newcommand{\bd}{\begin{displaymath}}
\newcommand{\ed}{\end{displaymath}}
\newcommand{\ba}{\begin{array}{ll}}
\newcommand{\ea}{\end{array}}
\newcommand{\baa}{\begin{eqnarray}}
\newcommand{\eaa}{\end{eqnarray}}
\newcommand{\baaa}{\begin{eqnarray*}}
\newcommand{\eaaa}{\end{eqnarray*}}   

\def\e{\varepsilon}

\def\defi{\stackrel{{\scriptscriptstyle \Delta}}{=}}

\def\a{\alpha}
\def\d{\delta}
\def\o{\omega}

\def\Y{{\cal Y}}

\def\w{\widehat}
\def\Ind{{\mathbb{I}}}

\def\sign{{\rm  sign\,}}

\def\R{{\bf R}}

\def\Z{{\cal Z}}
\def\ZZ{{\bf Z}}

\def\g{\gamma}
\def\C{{\bf C}}

\def\T{{\mathbb{T}}}
\def\TT{{\cal T}}
\def\ZZ{{\mathbb{Z}}}
\def\a{\alpha}

\def\ew{\left(e^{i\o}\right)}

\title{
On recoverability of discrete time  signals  from sparse observations  }
\author{Nikolai Dokuchaev }
\begin{document}
\def\break{}%
\def\brea{}
\def\breakk{}
\maketitle
{\xxxonly{ \let\thefootnote\relax\footnote{Submitted: May 8, 2019. Revised: April 13, 2020}}}
\begin{abstract}
     The paper investigates recoverability of  discrete time  signals represented by infinite sequences  from incomplte observations.
 It is shown that there exist  wide classes of signals that are everywhere dense in the space of square-summable signals
 and such that signals from these classes     feature
           robust linear recoverability of their finite traces under very mild restrictions on the location of
     the observed data. In particular, the case  arbitrarily sparse and non-periodic subsequences
     of observations are not excluded. 

\end{abstract}
\section{Introduction}
In general, possibility of recovery  of  a signal from a sample is usually associated with constraints
that ensures an uniqueness  of recovery for
the  classes of underlying signals such as restrictions on the spectrum support or signal sparsity.  Analysis of these
classes can also lead  to recovery methods for noise contaminated  signals;
  the corresponding recovery algorithms can be applied to the projections of the underlying processes on a recoverable class of signals.
 For example, in continuous time setting, band-limited functions  can be  recovered without error  from a discrete
sample taken with a sampling rate  that is at least twice the maximum frequency
present in the signal (the Nyquist critical rate).    This defines the class of recoverable functions and the set of observations required for the recovery.

 Clearly,  a process of  a general type cannot be approximated by
band-limited processes with a preselected band or by  processes with a sparse spectrum with a preselected degree of sparsity.  This leads to major limitations
for data compression and recovery. For example, consider data compression via approximation of a continuous time function by samples of band-limited functions. A closer  approximation would  require wider spectrum band  for these band-limited functions or more frequent sampling; respectively, a closer approximation leads to less efficient data compression.

Therefore, there is an important problem of finding wide enough classes of recoverable processes for different choices set of available observations.

For example, for continuous time signals with certain structure, it was found that the restrictions imposed by the Nyquist rate could be excessive for signal recovery; see e.g. \cite{BS,ME}.
In particular,  a sparse enough subsequence or a semi-infinite subsequence  can be removed from
an oversampling sequence \cite{F95,V87}.\index{(Vaidyanathan  (1987),  Ferreira (1995)).}
There is also a so-called  Papoulis approach \cite{Pa} allowing to reduce the  sampling rate with additional measurements at sampling points.
Very wide uniqueness classes of continuous time signals with unlimited spectrum support  were considered  in the framework of the approach based on the so-called Landau's criterion;
see. e.g.,  \cite{La,La2,OU}, and a recent literature review in \cite{OU}.

For finite discrete time  signals, some  paradigm changing results were obtained in \cite{CR1,Don06}
and consequent papers   \index{Cand\'es and Tao (2006),  Cand\'es et al (2006). } in the so-called  compressive sensing   setting.
This approach explores sparsity of signals, i.e. restrictions on the number of nonzero members  of the underlying finite sequences.

In general, there is a difference between the problem of uniqueness of recovery and the  problem of existence of a stable recovery algorithm.
 As was emphasized in \cite{La2},  the  uniqueness results   do not imply stable
  data recovery. For example,  any sampling below the Landau's  critical rate  cannot be stable. The Landau's rate mentioned here  is  a generalization
    of the critical Nyquist rate for the case of stable recovery, non-equidistant sampling and disconnected spectrum gaps.

The present paper considers infinite discrete time signals.
It is shown that there exist  wide classes of signals that are everywhere dense in the space of square-summable signals
 and such that signals from these classes     feature
           robust recoverability of finite traces under very mild restrictions on the location of
     the observed data  (Theorem \ref{ThM} below).
     In particular, the case  arbitrarily sparse and non-periodic subsequences
     of observations are not excluded. This result represent a generalization of  results  \cite{D17,D19} obtained for some special sets of observed points and special types of spectrum degeneracy.
     
     The paper presents the required predicting kernels explicitly via their Z-transforms. 

The paper is organized as following. Section \ref{SecDef} presents some definitions and preliminary results on predictability of sequences.  Section \ref{SecM} presents the main result.
 Section \ref{SecP} contains the proofs.
 Section \ref{SecD} presents some discussion.

\section{Definitions and background}\label{SecDef}
Let $\T\defi\{z\in\C:\ |z|=1\}$, and let $\ZZ$ be the set of all
integers. Let $\ZZ^-=\{t\in\ZZ:\ t\le 0\}$, and let  $\ZZ^+=\{t\in\ZZ:\ t> 0\}$.
\par
We denote by $\ell_r$ the set of all sequences
$x=\{x(t)\}\subset\C$, $t=0,\pm 1,\pm 2,...$, such that
$\|x\|_{\ell_r}=\left(\sum_{t=-\infty}^{\infty}|x(t)|^r\right)^{1/r}<+\infty$
for $r\in[1,\infty)$ or  $\|x\|_{\ell_\infty}=\sup_t|x(t)|<+\infty$
for $r=+\infty$.
\par
For  $x\in \ell_1$ or $x\in \ell_2$, we denote by $X=\Z x$ the
Z-transform  \baaa X(z)=\sum_{t=-\infty}^{\infty}x(t)z^{-t},\quad
z\in\C. \eaaa Respectively, the inverse $x=\Z^{-1}X$ is defined as
\baaa x(t)=\frac{1}{2\pi}\int_{-\pi}^\pi X\left(e^{i\o}\right)
e^{i\o t}d\o, \quad t=0,\pm 1,\pm 2,....\eaaa

We have that  $x\in \ell_2$ if and only if  $\|X\ew\|_{L_2(-\pi,\pi)}<+\infty$.   In addition, $\|x\|_{\ell_\infty}\le \|X\ew\|_{L_1(-\pi,\pi)}$.


For $\rho>0$, we denote $B_\rho(\ell_2)=\{x\in\ell_2:\ \|x\|_{\ell_2}\le \rho\}$.

For a finite set $S$, we denote by $|S|$ the number of its elements.

We denote by $\Ind$ is the indicator function.

\subsection*{The setting for the recovery problem}
Let disjoint subsets $\M$ and $\TT$ of $\ZZ$ be given, and let $\V\defi \ZZ\setminus(\M\cup \TT)$.

We are interested  in the problem of recovery  values
$\{x(t)\}_{t\in\TT}$  from  observations $\{x(s)\}_{s\in\M}$ for $x\in\ell_2$, possibly, in the presence of a contaminating noise.
We consider  linear estimates only.

\begin{definition}\label{def1} { Let $\X\subset \ell_2$ be a set of signals.
Consider  a problem of recovery $\{x(t)\}_{t\in\TT}$  from observations on $\M$ of
 noise contaminated sequences $x=\ww x+\xi$, where
 $\ww x\in\X$, and where $\xi\in\ell_2$ represents a noise.
 We say that  $\X$ allows finitely robust   $(\M,\TT)$-recovery
 if there  exists
  a sequence of mappings $h_n:\ZZ\times \ZZ\to \R$, $n=1,2,...$, such that $\sup_{t\in \TT} \|h_n(t,\cdot)\|_{\ell_2}<+\infty$ \comm{(a u nas? - OK)} and
 that, for any $r>0$ and $\e>0$,  and any finite set   $I\subset \TT$, there exists
$\rho>0$ and $N>0$  such that \baa &&\sup_{t\in \TT\cap I} |x(t)- \w x_n(t)|\le \e
\breakk \quad \forall \ww x\in\X \cap B_r(\ell_2), \ \eta\in B_\rho(\ell_2),\ n>N, \label{predu}\eaa
where  $x=\ww  x+\eta$ and
\baa
\w x_n(t)=\sum_{s\in \M,\ |s|\le N} h_n(t,s) x(s).
\label{wxx}\eaa}
\end{definition}
\comm{WRONG: there exists
$\rho>0$,  $\oo n$ and $N>0$  such that \baaa \sup_{t\in \TT\cap I} |x(t)- \w x_n(t)|\le \e\quad
\forall  x=\ww x+\eta,\quad \ww x\in\X,  \ n>\oo n,\ \eta\in B_\rho(\ell_2),  \label{predu}\eaaa}

\begin{proposition}\label{prop1}
If a set $\X$ features finitely robust  $(\M,\TT)$-recoverability, then this set
features finitely robust  $(\oo\M,\oo\TT)$-recoverability for any disjoint subsets $\oo\M$ and $\oo\TT$ of $\ZZ$
such that   $\M\subset \oo\M$ and  $\oo\TT\subset \TT$.
\end{proposition}

{\em Proof of Proposition \ref{prop1}}. Let  $\{h_n\}_{n=1}^\infty$ be such as required for $(\M,\TT)$-recoverability in   Definition \ref{def1}. Then the conditions of Definition \ref{def1} hold
 the pair  $(\oo\M,\oo\TT)$ if one selects  the corresponding functions  $\oo h_n(t,s)=h_n(t,s)\Ind_{\{s\in \M\}}$. $\Box$
\section{The main results}\label{SecM}
\begin{theorem}\label{ThM}
Assume that any  of the following conditions holds:
\begin{itemize}
\item[(A)]
$|\M\cap \ZZ^-|=+\infty$ and  $|\TT\cap \ZZ^-|<+\infty$.
\item[(B)]
 $|\M\cap \ZZ^+|=+\infty$ and  $|\TT\cap \ZZ^+|<+\infty$.
\item[(C)]
 $|\M\cap \ZZ^-|=+\infty$ and  $|\M\cap \ZZ^+|=+\infty$.
\end{itemize}
Then there exists a set of processes
$\mathcal{B}_{\M,\TT}\subset \ell_2$  that features finitely robust  $(\M,\TT)$-recoverability  and  such that, for any
$x\in\ell_2$ and  any $\e>0$, there exists $\w x\in \mathcal{B}_{\M,\TT}$ such that $\|\w x- x\|_{\ell_2}\le \e$ and  $\|\w x\|_{\ell_2}\le\|x\|_{\ell_2}$.
\end{theorem}
\vspace{4mm}
\begin{remark}{\rm
Theorem \ref{ThM} assumes that the set $\M$ is infinite but its statement actually implies that an estimate of a finite trace
of unknown values can be obtained  using a {\em finite} set of observations, even  contaminated by a noise, as described in Definition \ref{def1}. On the other hand,
to ensure that the recovery error does not exceed a preselected value,  one would need to include a sufficiently large number of observations.
}\end{remark} 
\begin{remark}{\rm
In the proof of Theprem \ref{ThM} below, the required predicting kernels are presented explicitly via their Z-transforms. 
}\end{remark}
\subsection*{Some examples where Theorem \ref{ThM} holds}
\begin{enumerate}
\item If
$\M=\{k,\ k\in\ZZ^-\}$ and $\TT=\ZZ^+$, then condition (A)  of Theorem \ref{ThM} is satisfied.
 \item
If $\M=\{k m,\ k\in\ZZ^-\}$ and $\TT=\ZZ^+$, where $m\in\ZZ^+$ is given, then condition (A)  of Theorem \ref{ThM} is satisfied.
 \item
If $\M=\{k^d,\ k\in\ZZ^-\}$ and $\TT=\ZZ^+$, where $m\in\ZZ^+$ and $d\in \ZZ^+$ are given, then condition (A)  of Theorem \ref{ThM} is satisfied.
\item If $\M=\{k^d,\ k\in\ZZ^+\}$ and  $\TT=\{k\in \ZZ,\ k<s\}$, where $d\in\ZZ^+$ and $s\in \ZZ$ are given then condition (B)  of Theorem \ref{ThM}  is satisfied.
\item If $\M=\{ m|k|^d\sign k,\ k\in\ZZ\}$ and  $\TT=\ZZ$, where $m\in\ZZ^+$ and $d\in\ZZ^+$,  then condition (C)  of Theorem \ref{ThM}  is satisfied.
\end{enumerate}
In the cases (i)-(iv), the recovery problem is a predicting problem.
In the cases  (ii)-(v), the subsequence of observations can be arbitrarily sparse.
 In the cases (iii)-(v), the sequences of observations are non-periodic, and there are infinitely growing gaps between observations.

 \section{Proof of  Theorem \ref{ThM}}\label{SecP}
 \subsection*{Case A}
 Let us prove first that Theorem \ref{ThM} holds if  condition  (A) is satisfied.
 We refer it as Case (A).  Let
 $\t=-1+\min_{t\in\TT}t$.  By condition (A) of the theorem, $\t>-\infty$.

 By Proposition \ref{prop1}, it suffices to consider the case where
 $\M\subset\{t\in\ZZ:\ t\le \t\}$. Let us assume that this is the case.

Let the sequence $\{\tau(k)\}_{k\in\ZZ}$ be such that $\tau(k-1)<\tau(k)$ for all $k$,
 \baa
 \M=\{\tau(k)\}_{k=-\infty}^\t,\quad \tau(k)=k \quad \hbox{for}\quad k>\t.
 \label{tau}
 \eaa


Let us consider a mapping $f_\tau:\ell_2\to \ell_2$    such that  $y(k)=x(\tau(k))$ for all $k\in\ZZ$ for  $y=f_\tau(x(\cdot))$.

For $\d>0$, let $J(\d)\defi\{\o\in(-\pi,\pi]: \ |e^{i\o}-1|\le \d\}$.

Let us define  a mapping $g_\d:\ell_2\to \ell_2$    such that  $\w y=g_\d(y)=\Z^{-1}\w Y$, where
$\w Y\ew =\Y\ew \Ind_{\{\o\notin J(\d)\}}$ and $Y=\Z y$.

  Let $\B^y$ be the set of all  $\w y =g_\d(y)$ for all $y\in\ell_2$ and all $\d>0$.

 Let  $\M^y\defi \{k\in\ZZ,\ k\le \t\}$ and $\TT^y\defi \{k\in\ZZ,\ k>\t\}$.

{
Let us introduce kernels $h_n(t,s)$, $t,s\in\Z$,  defined explicitly as 
\baaa
h_n(t,\cdot)=\Z^{-1} H_{n,t}, 
\eaaa
where 
\baaa
 &&H_{n,t}(z)=(z V_n(z))^{t-\t},\quad \brea
 V_n(z)\defi 1-\exp\left[-\frac{\g_n}{z+\a_n}\right], \qquad  z\in\C.
\label{wK}
\eaaa
Here $\a_n=1-\g_n^{-\w r}$, where
$\w r>0$ is fixed and where $\g_n\to +\infty$.
 
The equations for $h_n$ are based on the predictors introduced in \cite{D12}; they represent a reformulation of the corresponding equations  in \cite{D19}
adjusted  for the special case where the spectrum degeneracy is located in a neighbourhood of $z=-1$. 

 By Lemma 2 from \cite{D19}, the set  $\B^y$ features finitely robust $(\M^y,\TT^y)$-recoverability in the sense of Definition \ref{def1} with some the  kernels $\{h_n\}_{n>0}$,
 such that the required estimate of $y_n(t)$ is \baaa
&&\w y_n(t)=\sum_{s\in \M^y, |s|\le N_y} h_n(t,s)y(s)\breakk=\sum_{s=-N_y}^\t
h_n(t,s)y(s)
\eaaa
Here $N_y>0$ is a sufficiently large number.

Since $y_n(t)=x_n(t)$ for $t>\t$
the corresponding estimate  $\w x_n(t)$ for $t>\t$  
 can be presented as    \baaa
&&\w x_n(t)=\w y_n(t)=\sum_{s\in \M,\ s\ge - N} 
h_n(t,\tau(s)) x(\tau(s)).
\label{wyy}\eaaa
Here    $N=-\tau(-N_y)$.
}

Let us define a mapping $p_\d:\ell_2\to \ell_2$    such that  $x=p_\d(\ww x)$ is defined such that
\baaa
 &x(\tau(k))=\w y_\d(k),\quad &\hbox{if}\quad k\le \t, \\
 & x(s)=\ww x(s)\quad  &\hbox{if either }\quad s>\t\quad \hbox{or}\quad s\notin \M,
 \eaaa
where $\w y_\d=g_\d(y)$ and $y=f_\tau(\ww x)$.

 We  construct the sought set $\B_{\M,\TT}$ as  the set of all  $x =p_\d(\ww x)$ for all $\ww x\in\ell_2$ and all $\d>0$.

   It follows from the definitions and from
 the established recoverability of the set   $\B^y$ that the set  $\B_{\M,\TT}$ features finitely robust $(\M,\TT)$-recoverability
 in the sense of Definition \ref{def1} such that the required estimate can be presented as
 \baaa
\w x_n(t)=\sum_{s\in \M,\ |s|\le N_1} h_n(t,\tau(s))x(s).
\label{wxxn}\eaaa
 \subsection*{Case B}

Similarly to the Case A, we obtain that condition (B) is sufficient to ensure that the statement of the theorem holds. For this, we can  just repeat  the  proof adjusted to the use of backward prediction. We refer it as Case (B).

 \subsection*{Case C}
Let us prove that condition  (C) is sufficient to ensure that the statement of the theorem holds.

Let $\M_\pm \defi \M \cap \ZZ^\pm$ and  $\TT_\pm \defi \TT \cap \ZZ^\pm$.
Further, for $\ww x\in\ell_2$, let $\tau_+$, $y_+$,  $ x_+$, and $\B_+^y$, be defined
 similarly to   $\tau$, $y$, $x=p_\d(\ww x)$, and $\B^y$, respectively, defined for Case (A) with $\t=0$. By the result obtained for the Case (A), it follows that
the class  $\B_{\M_+,\TT_-}$ features finitely robust  $(\M_+,\TT_-)$-recoverability, i.e., the conditions of Definition \ref{def1} hold,
with the estimates
\baaa
\w x_{n,+}(t)=\sum_{s\in \M_-, s\ge -N} h_{n,+}(t,\tau_+(s)) x_-(s),\quad t\in \TT_+.
\label{wxxn+}\eaaa
Here kernels $h_{n,+}^y$ are such as
 required in Definition \ref{def1}.

 Further, for $\ww x\in\ell_2$, let  $\tau_-$, $y_-$, $x_-$,  and $\B_-^y$, be defined
 similarly to $\tau$, $y$, $x=p_\d(\ww x)$, and  $\B^y$, respectively,  for Case  (B) with $\t=0$.
 By the theorem statement for  Case (B), it follows that
the class  $\B_{\M_+,\TT_-}$ features finitely robust $(\M_+,\TT_-)$-recoverability, i.e., the conditions of Definition \ref{def1} hold,
with the estimates
\baaa
\w x_{n,-}(t)=\sum_{s\in \M_+, s\le N} h_{n,-}(t,\tau_-(s)) x_-(s),\quad t\in \TT_-.
\label{wxxx-}\eaaa
Here   kernels $h_{n,-}^y$ are such as
 required in Definition \ref{def1}.

We  construct the sought set  $\B_{\M,\TT}$  as the class of processes $x\in \ell_2$ that can be represented as
 \baaa
  x_n(t)=x_{n,+}(t)\Ind_{\{t\in \M_+\}}+ x_{n,-}(t)\Ind_{\{t\in \M_-\}}
 \eaaa
for some $x_{n,+}(t)\in \B_{\M_+,\TT_-}$ and $x_{n,-}(t)\in \B_{\M_-,\TT_+}$.

 Clearly, an estimate (\ref{predu}) for a given $\e$  holds for sufficiently large $\oo n$ and small $\rho$, since
 similar estimates hold for $\sup_{t\in \M_\pm}| \w x_{n,\pm}(t)-x(t)|$.

 Let us show that $\w x_n$ can be represented via (\ref{wxx}) for with some choice of appropriate mappings $h_n:\ZZ\times\ZZ\to \R$.
By the definitions, it follows   that, for $t\in\TT$,
\baaa
 \w x_n(t)= \Ind_{\{t\in \M_-\}}\sum_{s\in \M_-,s\ge -N } h_{n,+}(t,\tau_+(s)) x_-(s)\brea+\Ind_{\{t\in \M_+\}}\sum_{s\in \M_+,s\le N} h_{n,-}(t,\tau_-(s)) x_-(s).
 \eaaa
 Hence
\baaa
 \w x_n(t) =\sum_{s\in \M, |s|\le N} h_{n}(t,s) x(s),
 \eaaa
 where
\baaa
&& h_n(t,s)\breakk= \Ind_{\{t\in \M_+\}}h_{n,+}(t,\tau_+(s))+\Ind_{\{t\in \M_-\}}h_{n,-}(t,\tau_-(s)).
 \eaaa
 This gives representation (\ref{wxx}).
This completes the proof for the Case (C) as well as the proof of Theorem \ref{ThM}. $\Box$

\section{Discussion}\label{SecD}
Theorem \ref{ThM} provides an existence result.
The conditions  on the choice of the sets $\M$ and $\TT$ imposed by Theorem \ref{ThM} are quite mild. There are no  restrictions on the sparsity of $\M$ or on the choices of  $r,I,\e,\d$ presented in Definition \ref{def1} and in the definition of $J(\d)$. For example, the set $\M$ can have arbitrarily located gaps, in particular,
it can have  periodic gaps as well as non-periodic  gaps. 

The proof  of Theorem \ref{ThM} provides explicitly a recovery algorithm based one a similar algorithm  \cite{D19} focused on a special case of 
periodic set $\M$; some problems for its numerical implementation  are
 outlined therein. Some experiments are described in \cite{D19}.
 In particular, if $\M$ is too sparse, or $I$ and $r$ are  too large, or $\e$ and $\d$ are to small, then
 the  corresponding predicting kernels
 will be too large and too heavy-tailed for implementation on standard computers.
 The  numerical examples  in \cite{D19} show   that an effective numerical  implementation  of this algorithm would require significant efforts.

There are other choices of the sets $\B_{\M,\TT}$ and of the recovery algorithms in the proof of Theorem \ref{ThM}.
The proof above uses
sets of band-limited processes with spectrum gaps $J(\d)$. Alternatively, they could  be replaced by processes
featuring spectrum degeneracy at a single point only, as is allowed in Lemma 2 \cite{D19}.
For some important special cases, other linear predicting kernels could be more effective. For example, \comm{the linear time-invariant  recovering operators suggested in \cite{D18xxx} would be  preferable in  the case where the set $\TT$ is finite. Probably, } the  recovering operators suggested in \cite{D17} would be  preferable in  the case where the set $\TT$ is finite and where the underlying processes belong to $\ell_1$. The linear recovery operators from \cite{D19} can be used in the case where the $\M$ is a periodic subsequence of $\ZZ$.
An approaches from the proofs of  \cite{F95,V87} could lead to a different  proof of  Theorem \ref{ThM}. However,
it is unclear if the  numerical feasibility can be improved by any of these possible modification.
\index{(Vaidyanathan  (1987),  Ferreira (1995)).}

We leave these questions  for the future research.
\index{\subsection*{Acknowledgments}  This work  was supported by Small Grants
Program of 
Faculty of Science and Engineering Research and Development Committee of Curtin University, Bentley.}


\begin{thebibliography}{100}
\bibitem{BS}
Butzer, P.L., and Stens, R. L.   (1992). Sampling theory for not necessarily band-limited functions: A historical overview. {\em SIAM Review} 34(1), 40--53.
\bibitem{CR1} Cand\'es, E., Romberg, J., and Tao, T. (2006). Robust uncertainty principles: Exact signal
reconstruction from highly incomplete frequency information.  \textit{IEEE Trans. Informat. Theory} 52(12), 52(2), 489--509.
\bibitem{Don06} Donoho, D.L. (2006). Compressed sensing. {\em IEEE Trans. on Inform. Theory} 52, 1289--1306.
\bibitem{D12} 
 Dokuchaev, N. (2012). Predictors for discrete time processes with
energy decay on higher frequencies. {\em IEEE
Transactions on Signal Processing} {\bf 60}, No. 11, 6027-6030.

\bibitem{D17} Dokuchaev, N. (2017).
On exact and optimal recovering  of missing values for sequences. {\em Signal Processing}  {\bf 135}, 81--86.
\comm{
\bibitem{D18xxx}
Dokuchaev, N. (2018). On recovery of signals with single point spectrum
degeneracy. arXiv:1809.09983.}
\bibitem{D19}
Dokuchaev, N. (2019). On recovery of discrete time signals from their periodic subsequences.
{\em Signal Processing} {\bf 162}  180--188.
\bibitem{F95}\index{[Ferreira (1995b)]}
 Ferreira, P. G. S. G. (1995).  Sampling series with an infinite number of unknown samples. In: {\em SampTA'95}, 268--271.
\bibitem{La}
Landau, H.J.  (1964). A sparse regular sequence of exponentials closed on large sets.{\em  Bull.
Amer. Math. Soc.} {\bf 70} 566--569.
\bibitem{La2}
Landau, H.J.  (1967). Sampling, data transmission, and the Nyquist rate. {\em Proc. IEEE} {\bf 55} (10), 1701-1706. 
\bibitem{ME} Mishali, M., Eldar, Y.C. (2011). Sub-Nyquist sampling: Bridging theory and practice. {\em IEEE Sign. Proc. Mag.} 28(6), 98--124.

\bibitem{OU}  Olevskii, A.M., Ulanovskii, A. (2016). {\em Functions with Disconnected Spectrum.} { Amer. Math. Soc.}, Univ. Lect. Ser. Vol. 46.

\bibitem{Pa} Papoulis, A. (1977). Generalized Sampling Expansion, {\em IEEE Trans. Circuits and Systems}  {\bf 24} (11), 652--654.
\bibitem{V87}
Vaidyanathan P. P. (1987). On predicting a band-limited
signal based on past sample values, {\em Proc. IEEE} {\bf 75} (8), 1125--1127. 
\bibitem{WV} Wu, Y., Verdu, S. (2010).  R\'enyi  Information Dimension: Fundamental Limits of Almost
Lossless Analog Compression, {\em IEEE Trans. on Inform. Theory} 56, 3721--3748.
\end{thebibliography}
\end{document}